\documentclass[useAMS,usenatbib]{mn2e}

\usepackage{caption}
\usepackage{natbib}
\usepackage{graphicx}
\usepackage{textcomp}
\usepackage{latexsym}
\usepackage{varioref}
\usepackage{xspace}
\usepackage{makeidx}
\usepackage{verbatim}
\usepackage{tabularx}
\usepackage{epstopdf}
\usepackage{amsmath}
\usepackage{array}
\usepackage{rotating}
\usepackage{hyperref}
\usepackage{float}

\def\Kepler{\textit{Kepler}}

\def\MAXI{\textit{MAXI}}
\def\K2{\textit{K2}}
\def\HERMES{\textit{HERMES}}

\title[Sco X-1 with \K2, \MAXI\ and \HERMES]
{Sco X-1 revisited with \Kepler, \MAXI\ and \HERMES: outflows, time-lags and echoes unveiled}
\author[S. Scaringi et al.]
{S. Scaringi$^{1}$\thanks{E-mail: simo@mpe.mpg.de}, T.~J. Maccarone$^{2}$,  R.~I. Hynes$^{3}$, E. K{\"o}rding$^{4}$, G. Ponti$^{1}$, C. Knigge$^{5}$, \newauthor C.~T. Britt$^{2}$ and H. van Winckel$^{6}$\\ 
$^{1}$Max Planck Institute f{\"u}r Extraterrestriche Physik, D-85748 Garching, Germany\\
$^{2}$Department of Physics, Texas Tech University, Box 41051, Lubbock, TX 79409-1051, USA\\
$^{3}$Louisiana State University, Department of Physics and Astronomy, Baton Rouge LA 70803-4001, USA\\
$^{4}$Department of Astrophysics/IMAPP, Radboud University Nijmegen, P.O. Box 9010, 6500 GL Nijmegen, The Netherlands\\
$^{5}$School of Physics and Astronomy, University of Southampton, Hampshire SO17 1BJ, United Kingdom\\
$^{6}$Instituut voor Sterrenkunde, KU Leuven, Celestijnenlaan 200D, B-3001 Heverlee, Belgium\\
}

\begin{document} 

\date{}

\pagerange{\pageref{firstpage}--\pageref{lastpage}} \pubyear{2015}

\maketitle

\label{firstpage}

\begin{abstract}
Sco X-1 has been the subject of many multi-wavelength studies in the past, being the brightest persistent extra-solar X-ray source ever observed. Here we revisit Sco X-1 with simultaneous short cadence \Kepler\ optical photometry and \MAXI\ X-ray photometry over a 78 day period, as well as optical spectroscopy obtained with \HERMES. We find Sco X-1 to be highly variable in all our datasets. The optical fluxes are clearly bimodal, implying the system can be found in two distinct optical states. These states are generally associated with the known flaring/normal branch X-ray states, although the flux distributions associated with these states overlap. Furthermore, we find that the optical power spectrum of Sco X-1 differs substantially between optical luminosity states. Additionally we find rms-flux relations in both optical states, but only find a linear relation during periods of low optical luminosity. The full optical/X-ray discrete correlation function displays a broad $\approx12.5$ hour optical lag. However during the normal branch phase the X-ray and optical fluxes are anti-correlated, whilst being correlated during the flaring branch. We also performed a Cepstrum analysis on the full \Kepler\ light curve to determine the presence of any echoes within the optical light curve alone. We find significant echo signals, consistent with the optical lags found using the discrete cross-correlation. We speculate that whilst some of the driving X-ray emission is reflected by the disk, some is absorbed and re-processed on the thermal timescale, giving rise to both the observed optical lags and optical echoes.  

\end{abstract}

\begin{keywords}
accretion, accretion discs - X-rays: binaries - X-rays: individual: Sco X-1
\end{keywords}

\section{Introduction}

Scorpius X-1 (hereafter Sco X-1) is a low-mass X-ray binary (LMXB), where a secondary main sequence star transfers material onto a magnetised neutron star (NS) via Roche-lobe overflow through an accretion disk. Sco X-1 was the first, persistently bright, extrasolar X-ray source to be discovered (\citealt{giacconi}). It lies at a distance of 2.8 kpc (\citealt{bradshaw99}), has an orbital period of 18.9 hours (\citealt{gottlieb75,cowley75,hynes12}), and contains a $0.4M_{\odot}$ M-type companion (\citealt{steeghs02}). Since its discovery, many studies have been pursued on Sco X-1, all revealing its highly variable nature across the electromagnetic spectrum (see e.g. \citealt{mook75,hertz92,dieters00}). In particular, Sco X-1 is a member of a group of objects called Z sources. The class name arises due to the fact that apparent Z-shaped tracks appear when soft X-ray colours are plotted against hard X-ray colours. Many multiwavelength studies have been pursued on Z-sources in the past, revealing how the mass accretion rate, ultraviolet line and continuum fluxes and optical brightness are all correlated with where the sources lie on the Z-track (\citealt{hertz92,hasinger90,vrtilek90,vrtilek91}). The top-most portion of the Z-track is referred to as the horizontal branch (HB), and systems have been shown to be rarely found in this state. As the sources progress downward, they enter the normal branch (NB), whilst during episodes of higher X-ray flux, they enter the bottom part of the Z-track called the flaring branch (FB). In the case of Sco X-1, the HB is far less pronounced than either the NB or FB. Additionally to X-ray colour changes the X-ray power spectral density (PSD) of Sco X-1 displays clear quasi-periodic oscillations (QPOs), which appear to increase in frequency as the source moves downwards along the NB (\citealt{vdKlis96}).  

One of the first models invoked to explain the Z-track phenomenology was that of \cite{priedhorsky86} and \cite{psaltis95}. In that model most of the X-ray emission originates from the NS magnetosphere, the hot central corona, and an extended corona through which material falls radially towards the NS. Variations within the Z-track then reflect variations in the mass accretion rate which varies between $0.5-1.1$ times the Eddington critical rate, $\dot{M}_{E}$. During low accretion rate episodes, Sco X-1 is found in the NB and moves downwards as $\dot{M}$ increases. This causes increased radiation pressure allowing material to pile-up close to the NS. The optical depth then increases, causing more X-ray photons to be absorbed and re-emitted as optical photons. This phase is associated with an X-ray to optical anti-correlation (\citealt{mcnamara03}). As the source transitions between the NB and the FB, the mass accretion rate exceeds the critical Eddington rate, causing the mass flow to become chaotic with material in the disk flowing both radially inwards and outwards. As material keeps building up in the outer corona the electron scattering opacity increases, redirecting some of the X-ray emitted photons onto the accretion disk. Contrary to the NB phase, the FB phase produces correlated X-ray and optical fluxes.

Recently, a different model has been proposed based on the idea of an extended accretion disc corona (ADC, \citealt{church95,church04,church12}). In this model $\dot{M}$ increases as the source moves upward along the NB track (the opposite of the \citealt{psaltis95} model). Although qualitatively different, this would still predict an X-ray/optical anti-correlation due to increased optical depths with higher accretion rates. Furthermore, movements along the FB away from the NB/FB vertex are also associated with increased $\dot{M}$, but here the flaring behaviour is associated to unstable nuclear burning on the NS surface (\citealt{church12}). The $\dot{M}$ increase along the FB is, however, only a property of Sco X-1 like sources: other systems like Cyg X-2 are instead thought to have constant $\dot{M}$ along the FB. Thus, given that most of the X-ray radiation is released from the NS rather than the disc, we might also expect that this scenario would produce correlated X-ray and optical fluxes as in the previous model. Furthermore, as movements along both FB and NB branches are associated with increased mass-transfer rates (and thus increased radiation pressure), the model predicts that both should display jet-launching. Indeed both branches have been shown to display radio emission, although the NB is more radio-loud than the FB (\citealt{hjellming90}). 

\cite{bradshaw07} have studied the correlations between the X-ray spectral characteristics and QPOs in Sco X-1. They found a strong correlation between the kHz QPOs and the spectral power law index and interpret this as being due to changes in the geometrical configuration of the corona. The same study shows how the equivalent width (EW) of the Iron $K_\alpha$ emission line is anti-correlated with spectral hardness. The increase in EW implies a stronger accretion disk wind, assuming the Iron $K_{\alpha}$ line is generated in the wind itself.  

With the aim of testing and refining the current models for Sco X-1, we revisit this system with simultaneous data obtained from the \K2\ mission and the \MAXI\ instrument, augmented by ground based optical spectroscopy obtained with \HERMES. Our data reduction procedures are described in Section \ref{sec:obs}, whilst our results are presented in Section \ref{sec:results}. Our discussion, together with interpretation of our results, is discussed in Section \ref{sec:discussion}, whilst our conclusions are drawn in Section \ref{sec:conclusion}.

\section{Data reduction and analysis}\label{sec:obs}
In this section we present the three datasets used in this work (\K2, \MAXI, \HERMES), and describe any relevant data reduction procedures employed. 

\subsection{K2 light curve}

The \textit{NASA} \Kepler\ mission was successfully launched on March 7 2009. It continuously pointed towards the same 116 square degree field-of-view (FOV) during the entire mission, and obtained light curves with a 58.8 second cadence (short cadence, SC) or with a 30 minute cadence (long cadence, LC) during every quarter (Q) of observations (\citealt{borucki}). In May 2013, \Kepler\ entered a prolonged safe mode due to a second reaction wheel failure. Since then the mission has been re-purposed as the \K2\ mission, covering new areas of the sky with the same SC and LC modes and the same FOV size. \K2\ thus stares for 3 months at the same field before moving onto the next one. Although \K2\ is not able to obtain year-long light curves as the original \Kepler\ mission, it does allow to study the timing properties of a wider range of astrophysical objects as the mission sweeps through the so-called \Kepler\ ecliptic plane.

\begin{figure}
\includegraphics[width=0.45\textwidth]{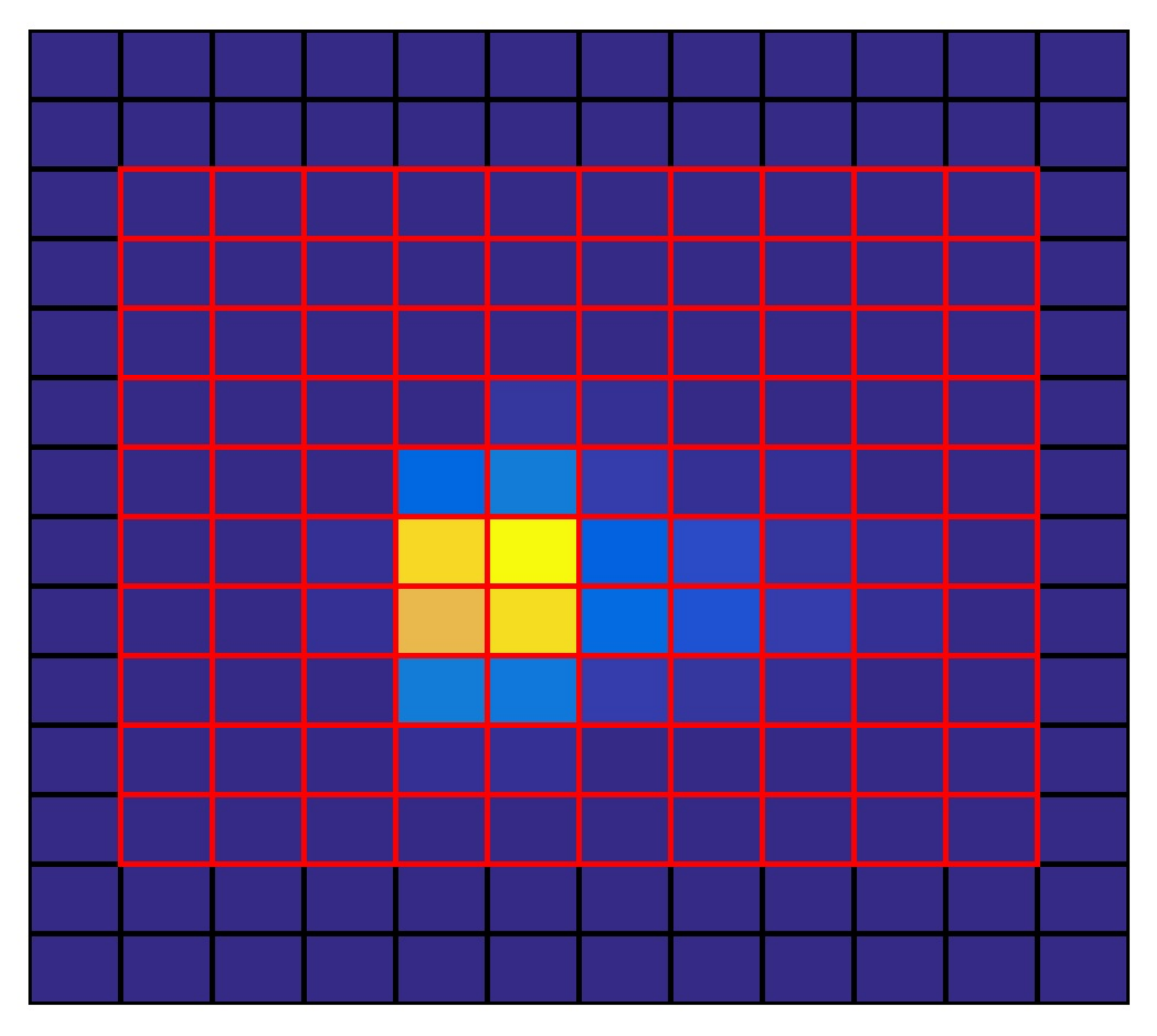}
\caption{\K2 average target pixel image for Sco X-1. The target mask pixels are marked in red, whilst background mask pixels are marked in black.}
\label{fig:mask}
\end{figure}

\begin{figure*}
\includegraphics[width=1\textwidth]{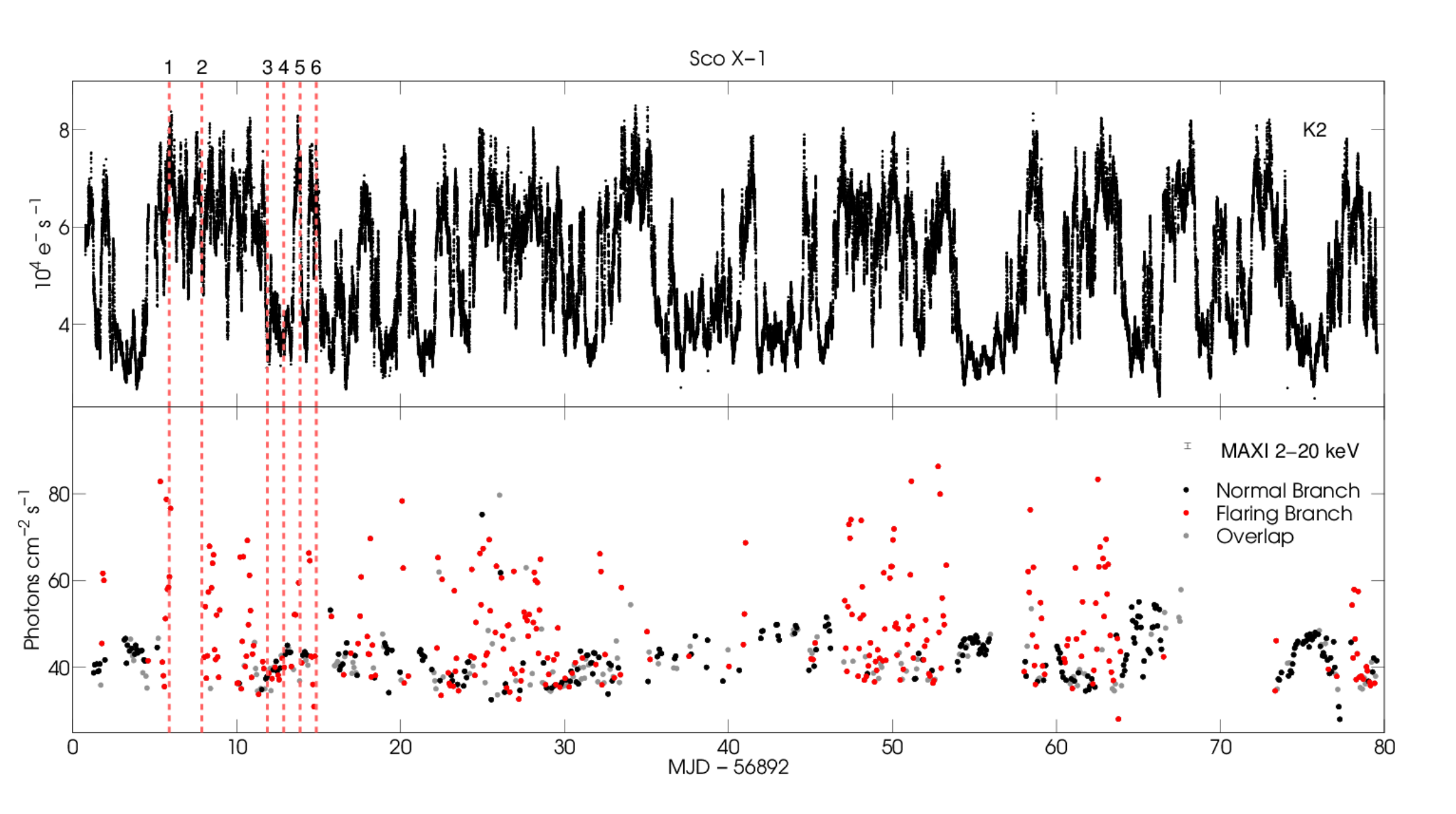}
\caption{Top panel: \K2 Sco X-1 light curve. The system was observed for over 78 days at 58.8 second cadence. The units on the y-axis are electrons/second, and can be converted to \Kepler\ magnitudes $Kp$ using the conversion found in the \Kepler\ Instrument Handbook. Bottom panel: 2-20 keV \MAXI\ light curve during the same period. Data points are colour coded using the NB/FB decomposition shown in Fig. \ref{fig:PCA}. Typical \MAXI\ error bar is shown on the top-right. The vertical dashed lines mark the times of our \HERMES\ observations.}
\label{fig:LC}
\end{figure*}

\begin{figure}
\includegraphics[width=0.45\textwidth]{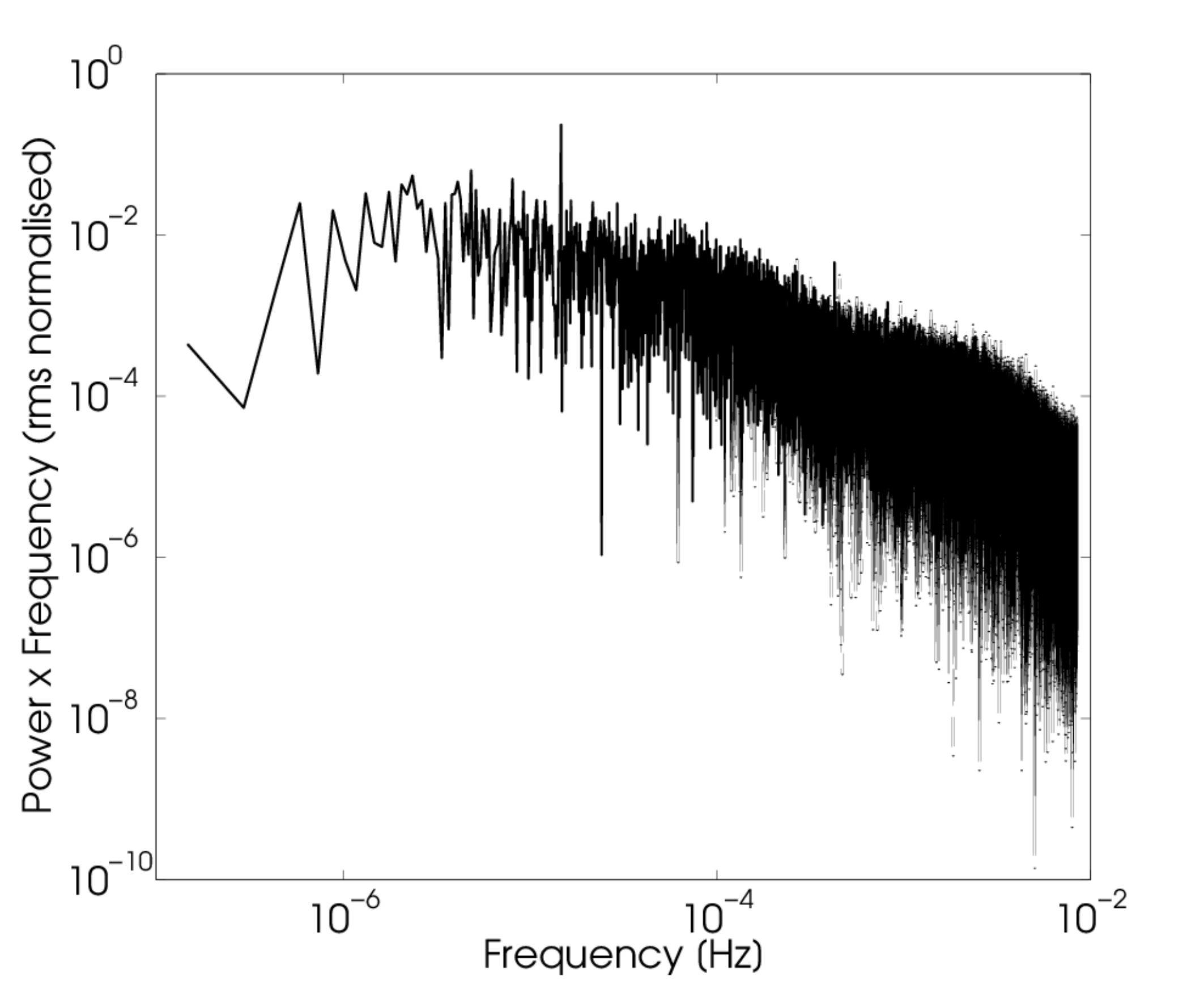}
\caption{Fourier Transform of the full 78.8 days SC \K2\ data for Sco X-1, showing a clear peak at the orbital period of 18.9 hours.}
\label{fig:FFT}
\end{figure}

Sco X-1 was observed by the \K2\ mission during Campaign 2 between August 23 and November 13 2014 (78.8 days), and is the first ever LMXB to be observed with \Kepler. Sco X-1 has a registered \Kepler\ magnitude ($Kp$) of $12.4$ in the \K2\ Ecliptic Plane Input Catalog (EPIC). Here we analyse SC data (cadence of 58.8 seconds) obtained from the Mikulski Archive for Space Telescope (MAST) archive\footnote{\url{http://http://archive.stsci.edu/k2/}}. The data is provided in raw format, consisting of target pixel data. For each 58.8 second exposure we thus have a 14 $\times$ 12 pixel image centered on the target. Sco X-1 was near the edge of module 15.3. Although this module is not known to be affected by Moire Pattern Drift (MPD) noise the target point spread function (PSF) is asymmetric as it lies at the edge of one of the outermost modules. As no other sources were present within the 14 $\times$ 12 pixel images, we created the light curve by manually defining a large target mask as well as a background mask. A large target mask is required due to occasional small scale jittering of the spacecraft, resulting in the target moving slightly from its nominal position. Fig. \ref{fig:mask} shows the average target pixel image obtained from 115,679 individual target images. We removed 1,919 observations because of bad quality due to occasional spacecraft rolls or due to cosmic rays. Fig. \ref{fig:mask} also shows in red the target mask and in black the background mask. We produce the lightcurve by summing together all target pixels for each exposure, and subtract the average background obtained from the background pixel mask. The obtained light curve is shown in the top panel of Fig. \ref{fig:LC}. We believe this is the best ever optical light curve obtained for an LMXB in terms of timespan, cadence and photometric quality. A Fourier transform of the whole light curve reveals the orbital period at 18.9 hours (see Fig. \ref{fig:FFT}).

\subsection{MAXI light curve}

\begin{figure}
\includegraphics[width=0.5\textwidth]{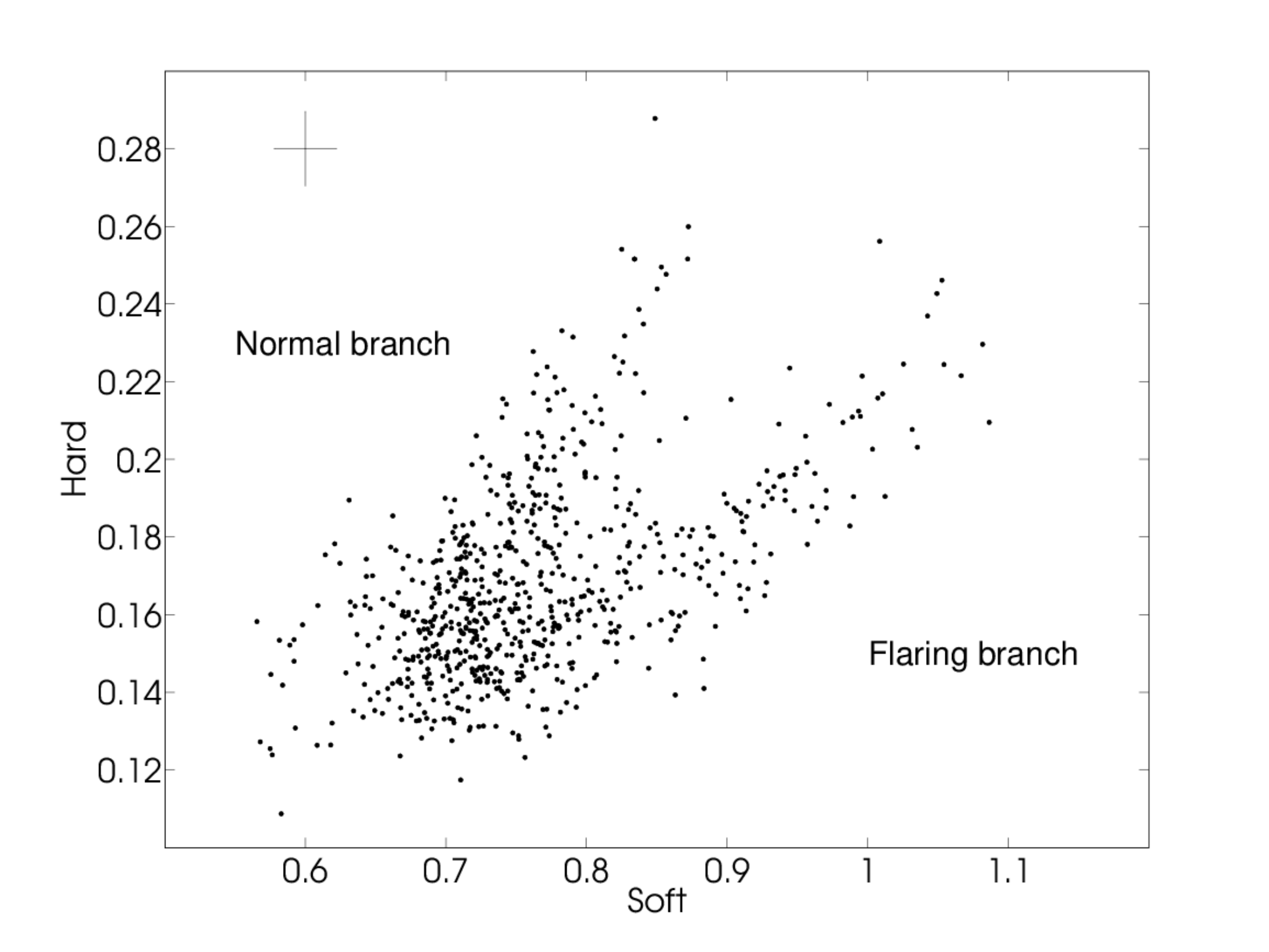}
\caption{Sco X-1 colour-colour X-ray diagram obtained from \MAXI\ data. The soft colour is defined as the ratio of 4-10 keV / 2-4 keV, whilst the hard colour is defined as the ratio of 10-20 keV / 4-10 keV. Typical error bar shown on the top-left.}
\label{fig:colcol}
\end{figure}

Since the \MAXI\ (Monitor of All-sky X-ray Image; \citealt{MAXI}) experiment on board the International Space Station (ISS) started in 2009 August, the GSC (Gas Slit Camera; \citealt{GSC1,GSC2}), one of the two MAXI detectors, has been scanning almost the whole sky every 92-minute orbital cycle in the 2-30 keV band. Here we use the \MAXI\ light curve of Sco X-1 during the period when the system was also observed by the \K2\ mission. We obtained the light curve through the \MAXI\ website\footnote{\url{http://maxi.riken.jp}}, which provides reduced light curves in the three energy ranges 2-4 keV, 4-10 keV and 10-20 keV. The combined light curve is shown in the bottom panel of Fig. \ref{fig:LC}. The corresponding X-ray colour-colour diagram is shown in Fig. \ref{fig:colcol}, clearly displaying the NB/FB dichotomy.

\subsection{Mercator/HERMES spectra}
\begin{figure*}
\includegraphics[width=1\textwidth]{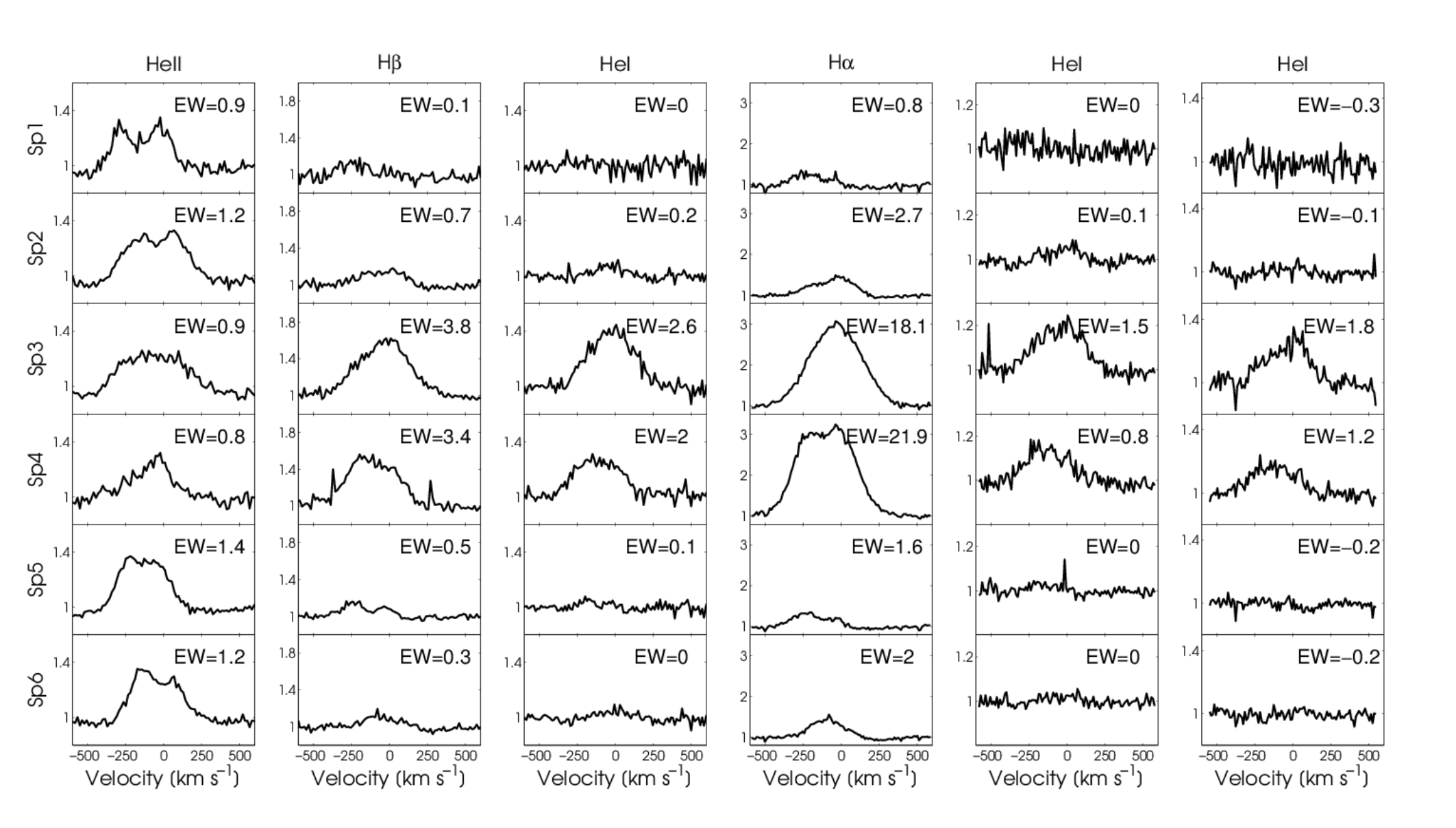}
\caption{\HERMES\ spectra normalised to continuum. Each row is one observation taken at times marked in Fig.\ref{fig:LC} with vertical dashed red lines. Columns are different zooms around different emission lines. Each panel also shows the measured equivalent width of the lines. Note the increased line equivalent width when Sco X-1 was in its Normal branch state (Spectra 3 \& 4).}
\label{fig:spec}
\end{figure*}

We observed Sco X-1 with the \HERMES\ spectrograph (\citealt{HERMES}), mounted on the 1.2 m Mercator Telescope at La Palma, Canary Islands, Spain. This highly efficient echelle spectrograph has a resolving power of $R=86000$ over the range $3800$\AA\ to $9000$\AA. Each spectrum was obtained from a 45 minute exposure. The raw spectra were reduced with the instrument-specific pipeline, but is not flux-calibrated. In total we obtained 6 spectra at the times marked with the vertical dashed lines in Fig. \ref{fig:LC}. Here we specifically examine the evolution of the 6 emission lines H$\alpha$ 6563 \AA, H$\beta$ 4861 \AA, HeI 5876 \AA, HeI 6678 \AA, HeI 7065 \AA\ and HeII 4685 \AA. To do this we normalise each line with an estimate of the neighbouring continuum. For each line, we iteratively fit a first degree polynomial around $40$\AA\ of the desired line centre using sigma-clipping. We found this to converge after about 5 iterations. In cases where no evident emission line was present, the iterative sigma-clipping procedure excluded $\approx3\%$ of data, whilst in cases where clear emission lines were present it excluded $\approx15\%$. The resulting normalised line profiles are shown in Fig. \ref{fig:spec}, together with their measured equivalent widths.

\section{Results}\label{sec:results}

In this Section we investigate the properties of Sco X-1 in the context of its NB and FB track dichotomy. We will determine what method best separates the 2 tracks using the \K2\ and \MAXI\ datasets and present average PSDs for the low and high optical luminosity states. This section additionally presents the results of cross-correlating the \MAXI\ and \K2\ light curves, the Cepstrum analysis to detect possible echoes in the \K2\ light curve, as well as our optical spectroscopy campaign results.

\subsection{Normal and Flaring branch properties}

\begin{figure*}
\includegraphics[width=1\textwidth]{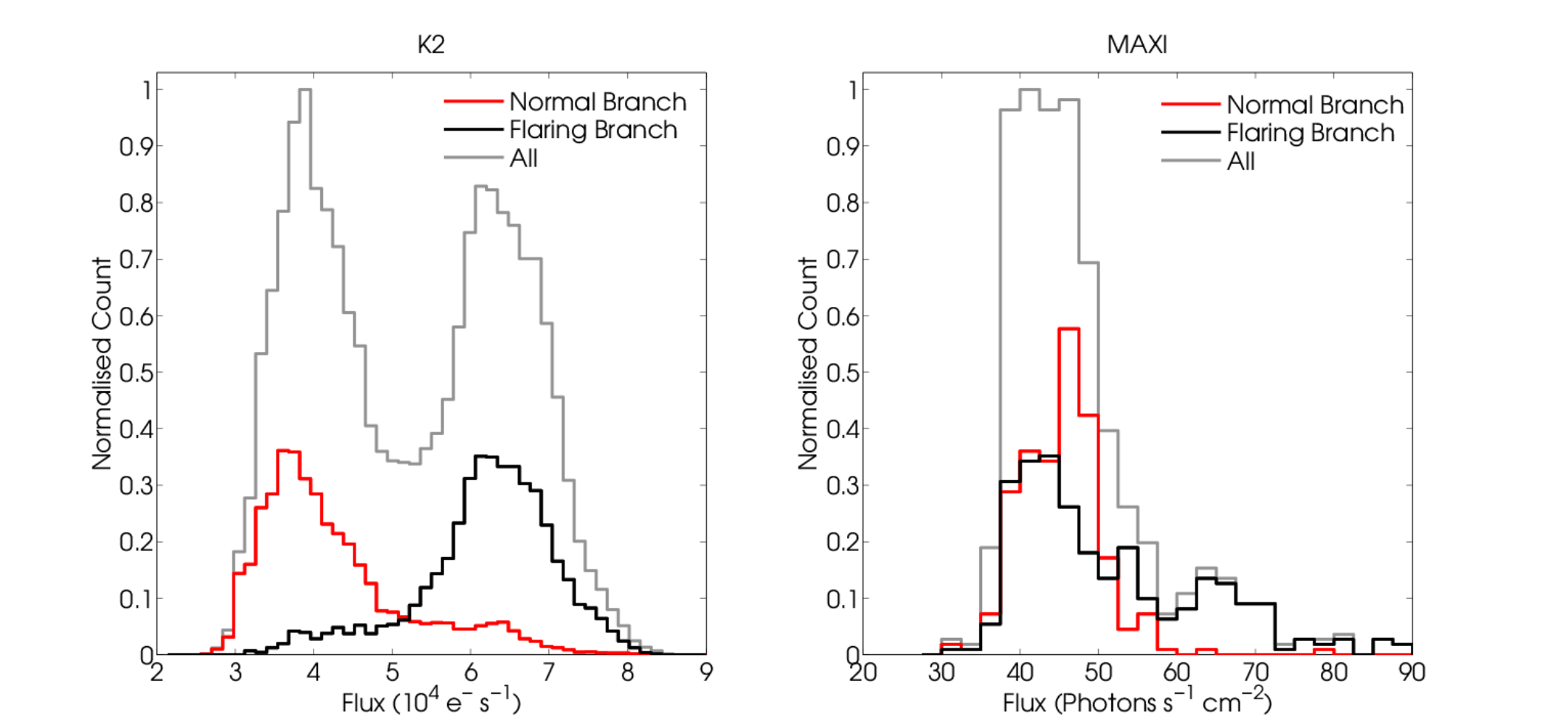}
\caption{Left panel: \K2\ flux distribution decomposed using PCA as applied to X-ray colours. Note that due to the conservative range employed the Flaring and Normal branch, data points are not all recovered. Right panel: \MAXI\ flux distribution decomposed using PCA as applied to the corresponding X-ray colours.}
\label{fig:fluxDist}
\end{figure*}

\begin{figure}
\includegraphics[width=0.5\textwidth]{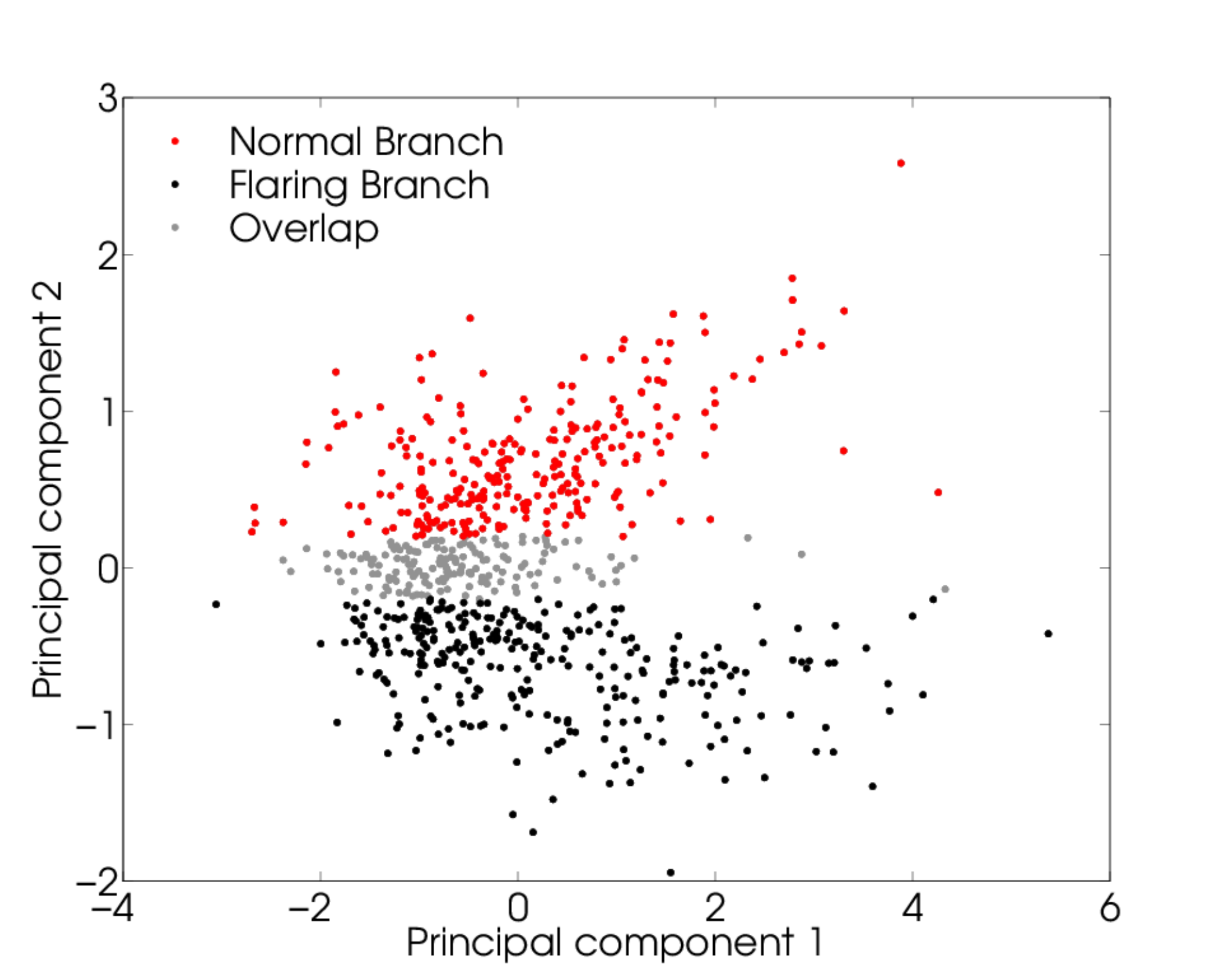}
\caption{Projected Hard and Soft X-ray colours using principal component analysis. We define conservative regions for the Flaring and Normal branch by selecting \MAXI\ observations that lie 0.2 above/below principal component 2. }
\label{fig:PCA}
\end{figure}

Fig. \ref{fig:fluxDist} shows the optical flux distribution obtained from the \K2\ light curve (left panel, gray line) and the X-ray flux distribution obtained from the \MAXI\ light curve (right panel, gray line). The \K2\ flux distribution is clearly bimodal, and possibly the \MAXI\ fluxes as well. From the flux distributions alone it is not clear whether the bi-modality is caused by the NB/FB dichotomy. To determine whether this is the case we performed principal component analysis (PCA) decomposition on the \MAXI\ X-ray colours shown in Fig. \ref{fig:colcol} in order to determine which observations belong to which Z-track branch. Fig. \ref{fig:PCA} shows the obtained PCA projection. By using principal component 2 ($PC2$) we can clearly separate the NB from the FB. We employ conservative ranges to exclude transition data points and select FB \MAXI\ observations as having $PC2>0.2$ and NB \MAXI\ observations as having $PC2<-0.2$. The resulting X-ray flux distributions for both the NB and FB obtained from our selected points are shown in Fig. \ref{fig:fluxDist}. Although NB X-ray fluxes seem to only populate the low flux distribution in Fig. \ref{fig:fluxDist}, the FB fluxes seem to populate both distributions. Thus X-ray fluxes alone are not a good discriminator between NB/FB tracks. We then located all \K2 observations that lie within $\pm45$ minutes from the \MAXI\ observations (\MAXI\ data points have a cadence of one ISS orbit, or 1.5 hours). The resulting FB and NB flux decomposition from our PCA projection is also shown in Fig. \ref{fig:fluxDist}. It is clear that in the absence of X-ray colour information, optical fluxes are better than X-ray fluxes in distinguishing between NB and FB. However we point out that optical fluxes alone are also not good indicators of the NB/FB dichotomy, given the long tails displayed in Fig. \ref{fig:fluxDist}.

\begin{figure}
\includegraphics[width=0.5\textwidth]{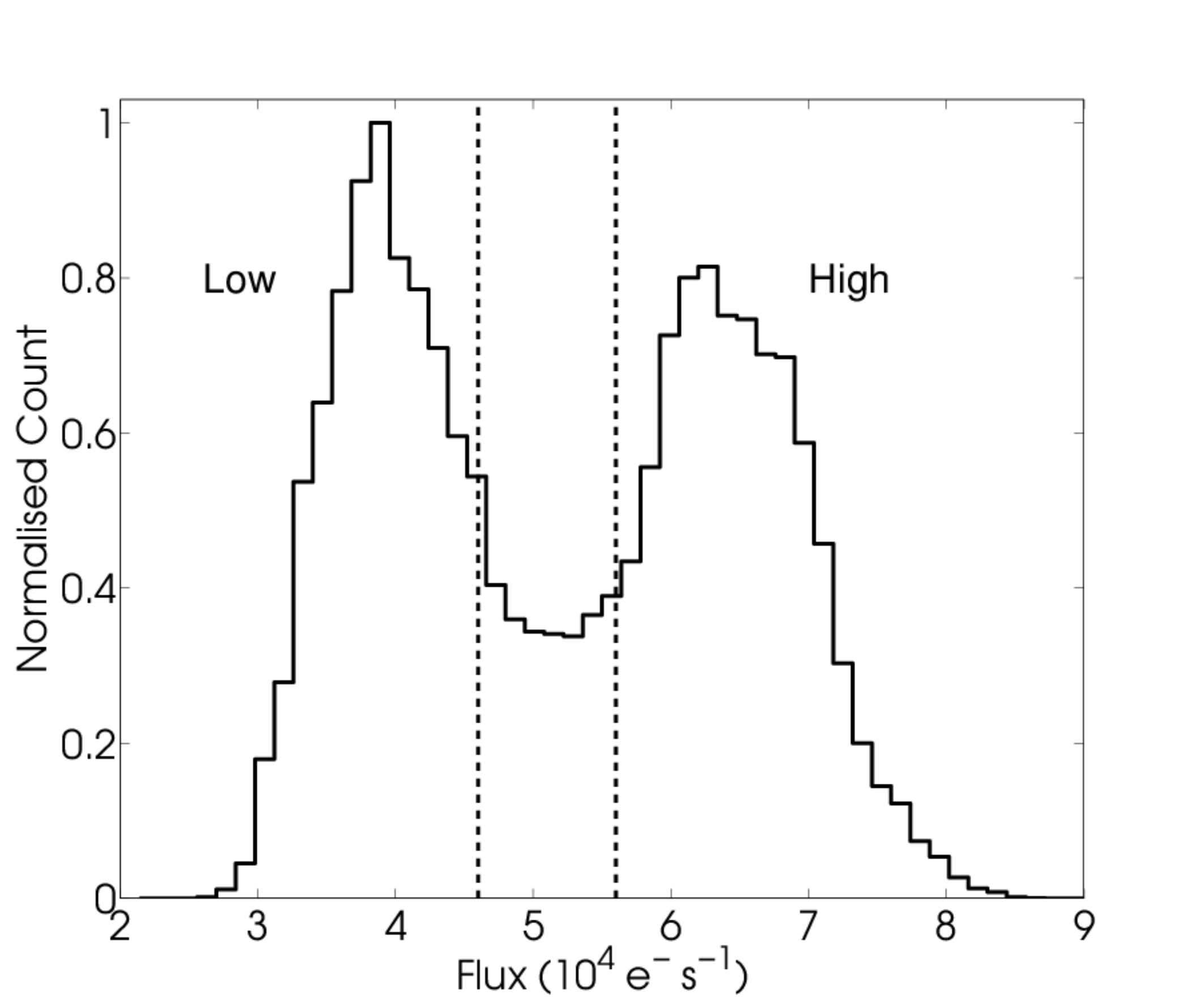}
\caption{Sco X-1 optical flux distribution obtained with \K2. The vertical lines correspond to conservative boundaries for low and high optical luminosity states determined by eye.}
\label{fig:fluxK2}
\end{figure}

\begin{figure}
\includegraphics[width=0.5\textwidth]{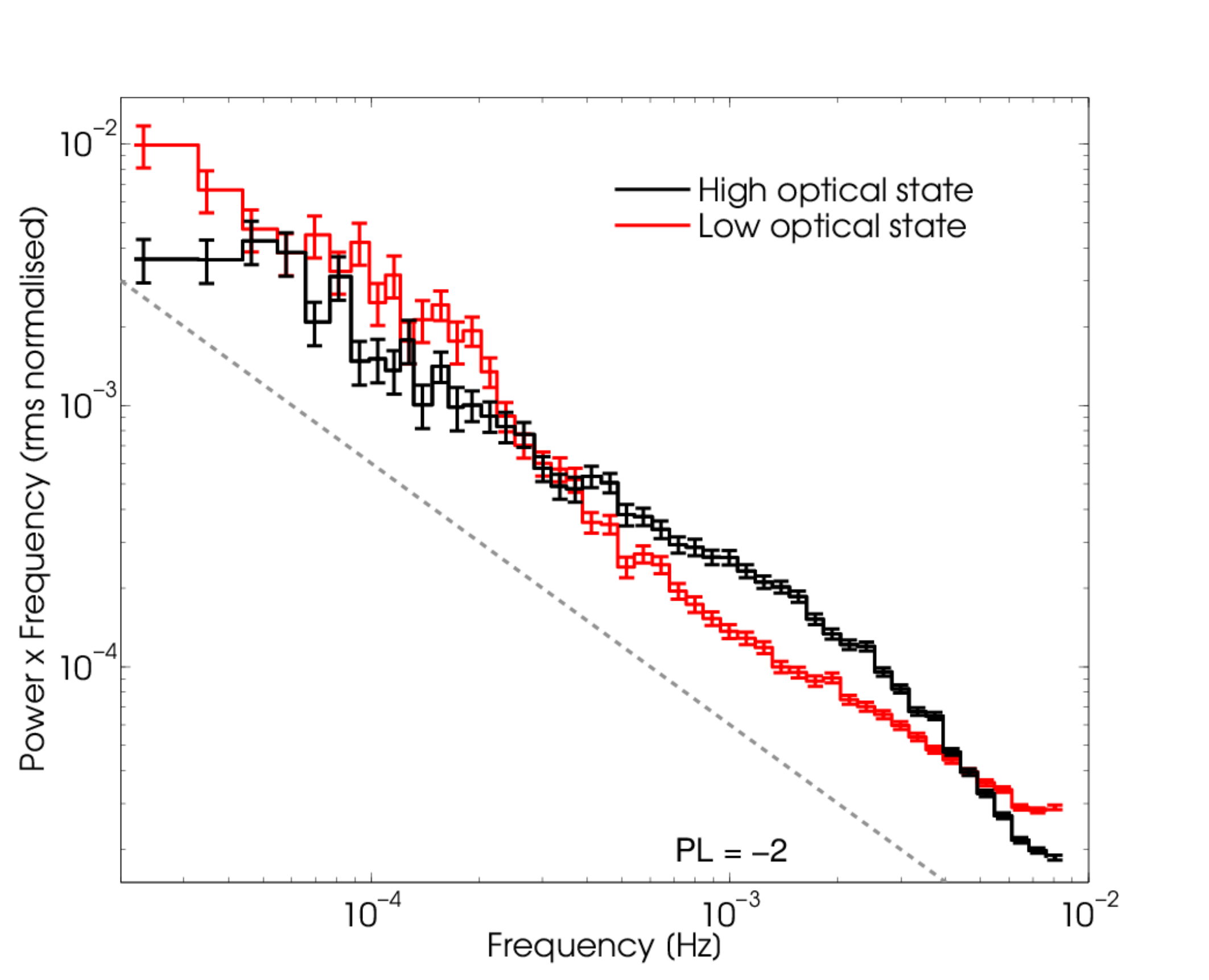}
\caption{PSDs obtained from the low and high optical luminosity intervals defined from the flux distribution shown in Fig.\ref{fig:fluxK2}. The dashed gray line shows a $-2$ power-law for comparison.}
\label{fig:PSDs}
\end{figure}

Given the optical flux bimodality, we decided to produce broad-band PSDs from the \K2 light curve for different flux levels. We selected conservative ranges to avoid the transition region, and chose to call low optical luminosity observations all data lying below $4.6\times10^4$ e$^-$ s$^{-1}$ and high optical luminosity those lying above $5.6\times10^4$ e$^-$ s$^{-1}$. These limits are marked with vertical lines in Fig. \ref{fig:fluxK2}. We then computed 1-day, non-overlapping PSDs (78 of them), and produced a low-luminosity PSD by averaging 30 of them whose mean fluxes lay within the optical flux range shown in Fig. \ref{fig:fluxK2}. The high optical luminosity PSD was similarly obtained by averaging 28 PSDs. The logarithmically binned result of this procedure is shown in Fig. \ref{fig:PSDs}. At first look it is already clear that the PSD shapes differ substantially between the low and high optical luminosity states. We include in Fig. \ref{fig:PSDs} a dashed gray line showing a power-law with a $-2$ gradient for comparison. During both high and low optical luminosity states the PSDs appear to follow a general $-2$ power-law shape. At low optical luminosities the PSD possibly displays an additional noise component around $\approx2\times10^{-4}$ Hz. On the other hand during high optical luminosities a clear broad noise component is observed in addition to the general $\approx-2$ power-law slope, peaking around $\approx10^{-3}$ Hz. The flattening observed at the highest frequencies is probably due to instrumental noise. It is clear that the PSD phenomenology is quite complex and differs substantially between states. A detailed modelling in terms of power laws and Lorentzian components is beyond the scope of this work and will be addressed in future. 

\begin{figure*}
\includegraphics[width=1\textwidth]{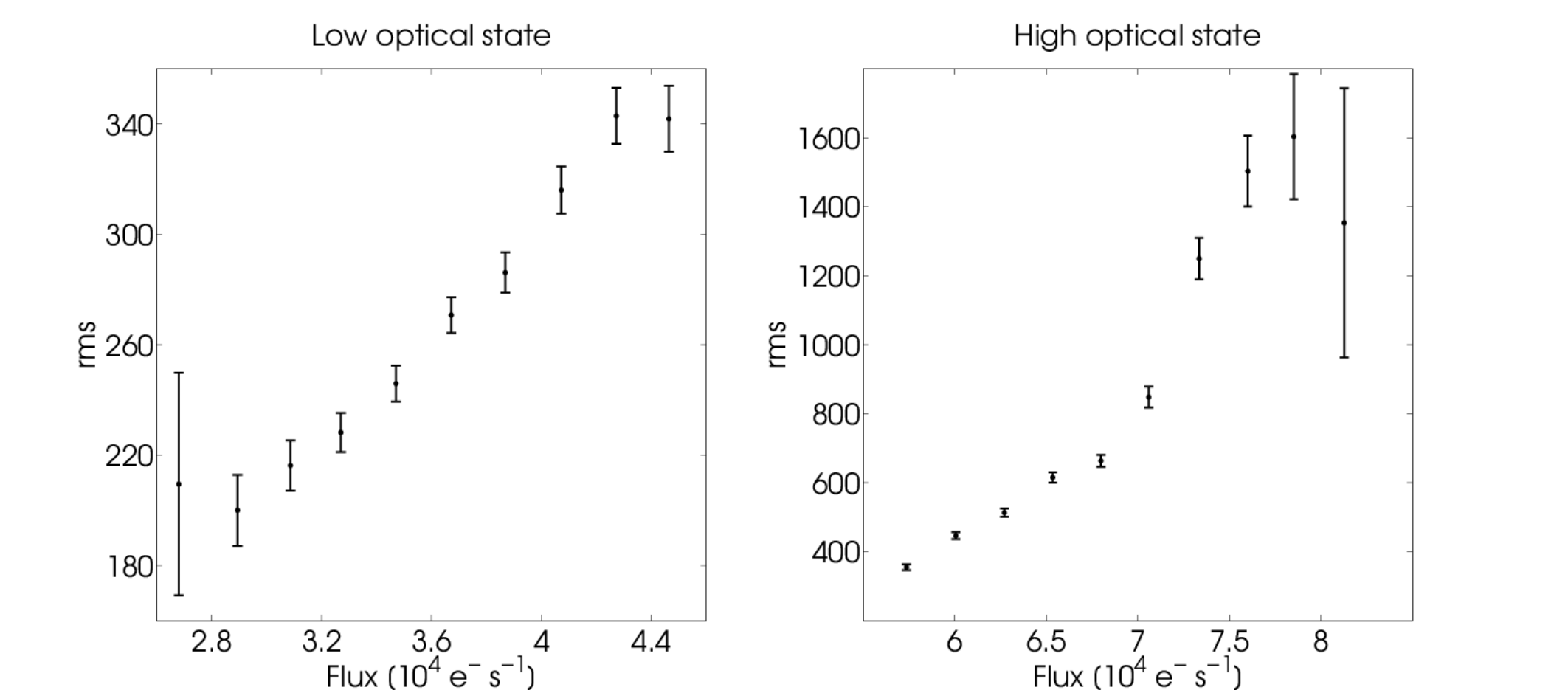}
\caption{rms-flux relations obtained for Sco X-1 during low and high optical luminosity states selected using the regions defined in Fig. \ref{fig:fluxK2}.}
\label{fig:rmsflux}
\end{figure*}

Using the optical flux limits defined in Fig. \ref{fig:fluxK2}, we are also able to produce root-mean-square vs. flux (rms-flux) relations for both the low and high optical luminosity states. This was done by computing mean fluxes and rms values for all non-overlapping 5-minute data segments, excluding any data gaps caused by the flux limit selections. Fig. \ref{fig:rmsflux} shows our binned results. In the low-luminosity state, an rms-flux relation consistent with linearity is obtained. In the high-luminosity state, it appears that linearity is broken at the highest fluxes. This could be suggestive that an additional flux distribution exists at the highest fluxes, thus making the light curve non-stationary and breaking the rms-flux linearity. We note that this could be connected to the trimodal flux distribution previously reported in Sco X-1 (\citealt{mook75,mcnamara03}) and/or the extra variability component observed during the high optical luminosity state as seen in Fig. \ref{fig:PSDs}. A similar property might also be observed in the NS XRB PSR J1023$+$0038 (\citealt{bogdanov15}), although we caution the reader that this particular system might have a higher accretion rate that Sco X-1, and thus any phenomenological similarity in the flux distributions might be coincidental. Nevertheless, it would be interesting to establish whether the rms-flux relation holds its linearity for this system or whether it breaks down at high fluxes as seen in Sco X-1. We also note that at the lowest fluxes the high optical luminosity rms-flux relation gradient is higher compared to the low luminosity one.

Fig. \ref{fig:spec} shows the 6 spectra obtained with the \HERMES\ spectrograph. Each column shows 6 different continuum normalised emission lines. The low ionisation emission lines of H$\alpha$, H$\beta$ and HeI qualitatively appear to have larger equivalent widths during periods of low optical luminosity (Spectra 3 \& 4) compared to the high luminosity spectra. On the other, hand the equivalent width of the higher ionisation HeII line does not appear to change by much. Furthermore, HeII appears to be double-peaked only during the high optical luminosity states phase. Possible explanations for this qualitative behaviour will be discussed in Section \ref{sec:discussion}.

\subsection{X-ray lags \& optical echoes}

\begin{figure}
\includegraphics[width=0.5\textwidth]{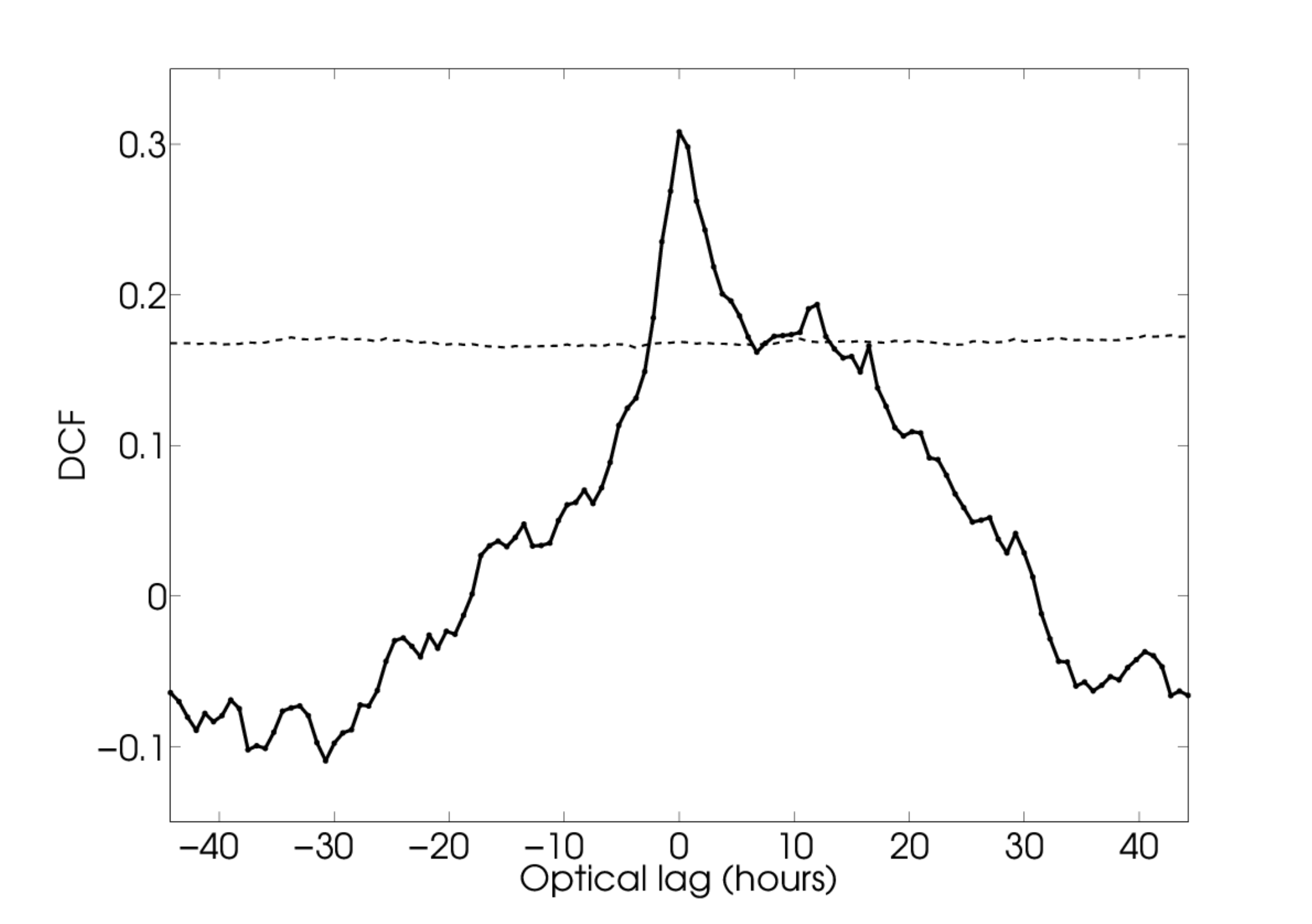}
\caption{Discrete cross-correlation between \MAXI\ and \K2\ data (solid line) and 3$\sigma$ detection level (dashed line).}
\label{fig:DCF}
\end{figure}

We computed the discrete cross-correlation function (DCF, \citealt{edelson88}) between the \K2\ and \MAXI\ 2-20 keV lightcurves using a time step of 45 minutes (half the \MAXI\ cadence). The results are shown in Fig. \ref{fig:DCF} where the x-axis is optical lag in units of hours. We further computed confidence levels by simulating $10^5$ lightcurves on the same \K2\ sampling pattern of the original data. Each light curve was simulated using the \cite{TK} algorithm. As we do not have an analytical model for the intrinsic PSD of Sco X-1 within the frequency range of interest, we produced the input PSD by first computing the full Fourier transform of the original \K2\ data and smoothing it with a 50-point moving average. The dashed line in Fig. \ref{fig:DCF} shows the obtained $3\sigma$ ($99.7\%$) contour level from our Monte-Carlo simulation. 

A clear significant zero-lag peak is evident. Given our timing resolution, this peak is consistent with the finding that the optical light curve lags the X-ray light curve by tens of seconds due to reprocessing from the secondary star and/or accretion disk (\citealt{teo07,brittPhD}). Additionally the DCF is very asymmetric, with a second broad significant bump around $12.5$ hours. This timescale might be consistent with the disk thermal timescale, and will be discussed further in Section \ref{sec:discussion}. 

\begin{figure*}
\includegraphics[width=1\textwidth]{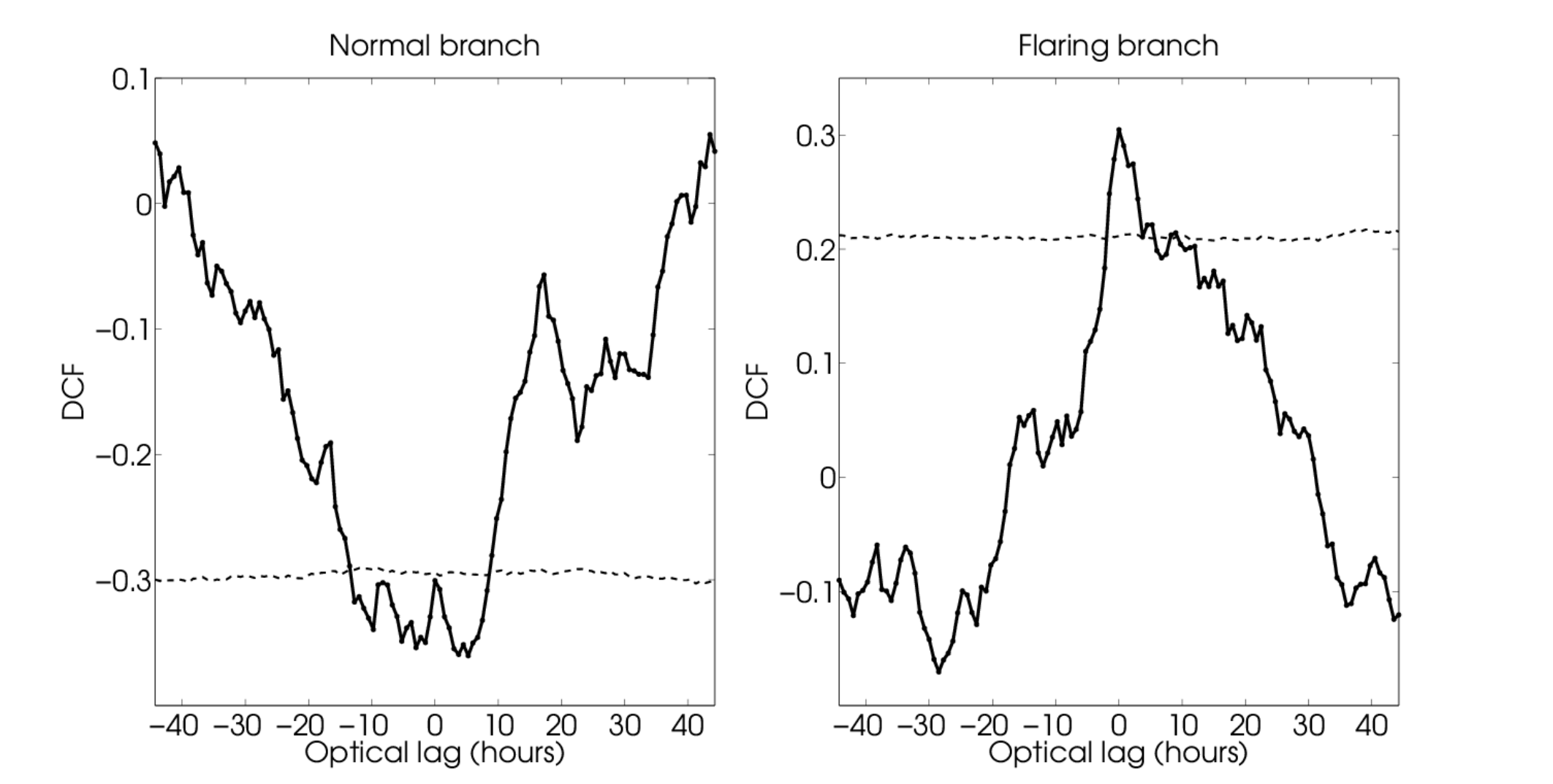}
\caption{Discrete cross-correlation between \MAXI\ and \K2\ data selected according to the PCA decomposition for normal and flaring branches (solid line, left and right respectively). The dashed lines mark the 3$\sigma$ detection levels.}
\label{fig:DCF2}
\end{figure*}

\begin{figure*}
\includegraphics[width=1\textwidth]{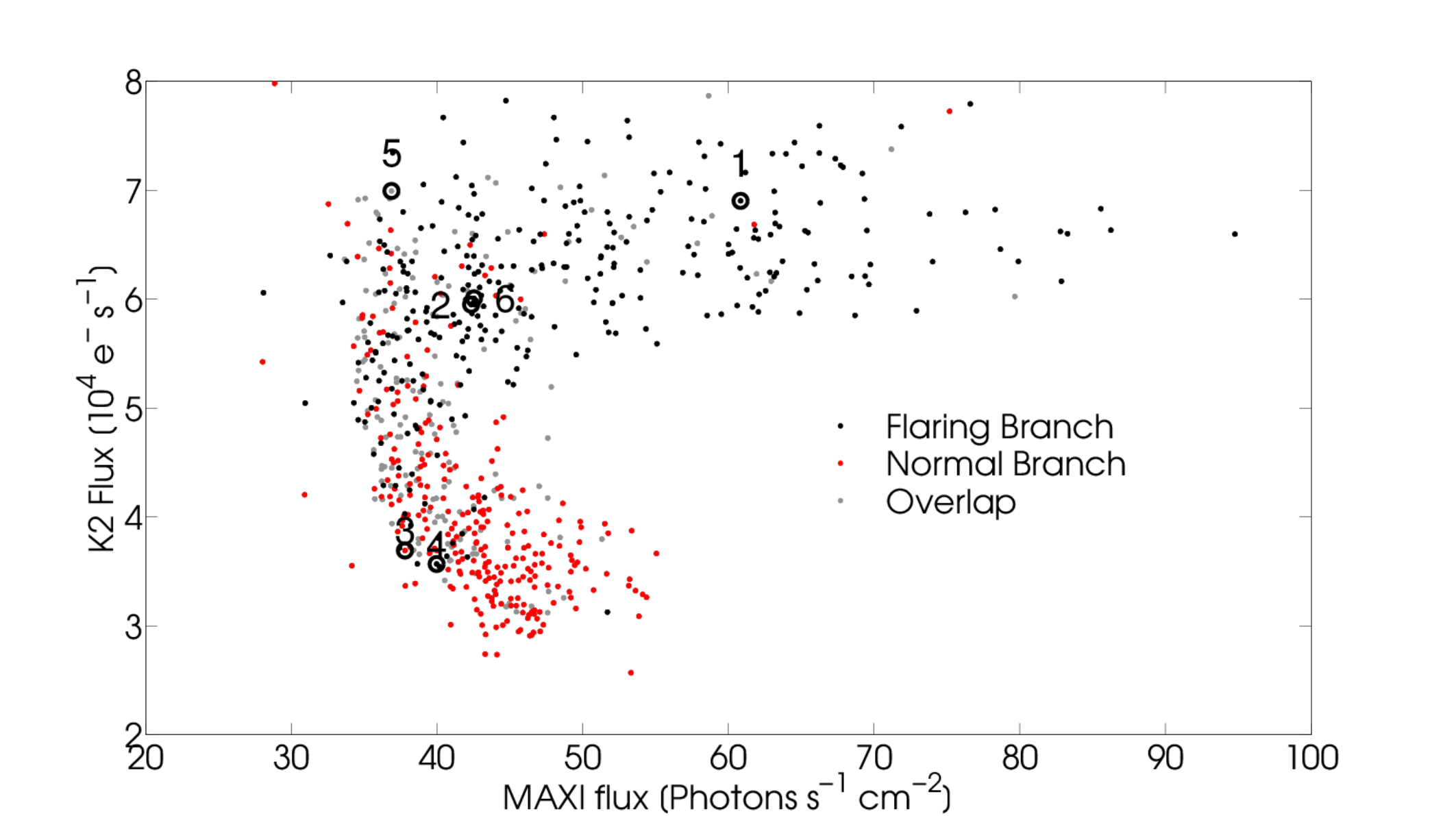}
\caption{\K2\ vs. \MAXI\ fluxes colour coded using the NB/FB PCA decomposition of Fig. \ref{fig:PCA}. Circles indicate the time when the 6 \HERMES\ spectra were obtained, numbered from 1 to 6 as in Fig. \ref{fig:spec}. }
\label{fig:K2vsMAXI}
\end{figure*}

Motivated by the different properties between the NB and the FB, as well as the idea that X-ray/optical light curves should be anti-correlated in the NB, we further computed the DCF for the two separate populations. We used the full \K2\ lightcurve for both, and selected \MAXI\ data segments using the PCA decomposition shown in Fig. \ref{fig:PCA}. The results of this are shown in Fig. \ref{fig:DCF2}, where the contour levels have been computed in the same way as for the whole dataset of Fig. \ref{fig:DCF}. Although slightly more scattered, the DCFs for the individual NB and FB show different qualitative behaviour. Similarly to the full dataset DCF, the FB DCF is asymmetric in the same direction. It displays a positive peak at zero time-lag, and although marginally significant, a potential secondary optical lag peak at $\approx4.5$ hours. During the NB the two light curves are anti-correlated and appear more symmetric than for the FB. This effect can also be noticed in the light curves themselves by close visual inspection (see e.g day $\approx75$ in Fig. \ref{fig:LC}). The correlation/anti-correlation properties between X-ray and optical fluxes have been already reported in the past (\citealt{mcnamara03}) using X-ray flux vs. optical flux plots, which we here confirm and characterise in more detail using the DCF. Fig. \ref{fig:K2vsMAXI} shows our version of X-ray vs. optical flux diagram, where we have additionally marked with circles the positions of when our optical spectra was taken. Simply from visual inspection it can be noticed that Spectrum 3 \& 4 (which have the strongest emission lines) have been taken at times where the optical flux was lowest, but where the X-ray flux is not so different as for the other spectra. This suggests that low ionisation line strength is more strongly correlated to optical flux and not X-ray fluxes or colours.

\begin{figure*}
\includegraphics[width=1\textwidth]{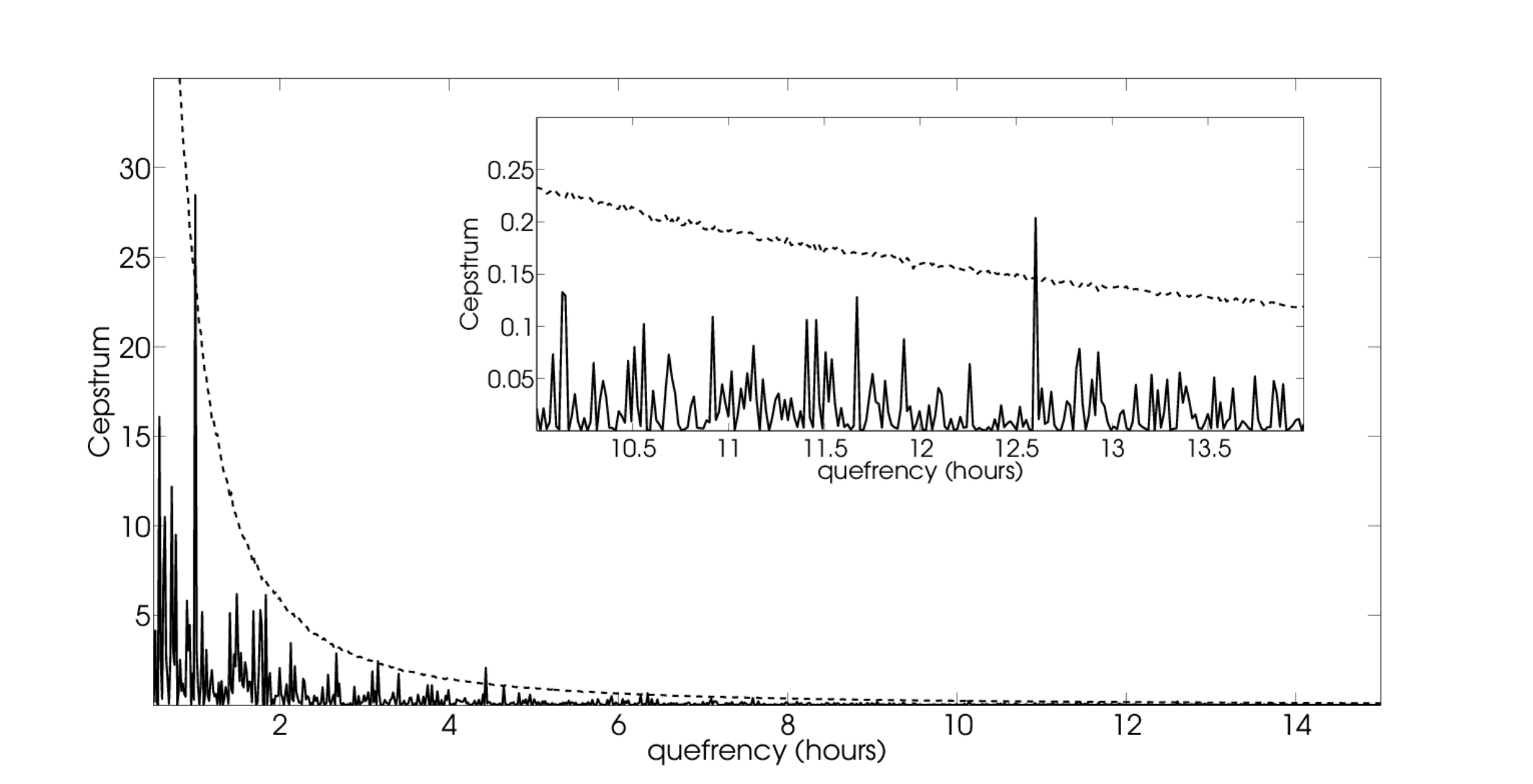}
\caption{Cepstrum obtained from the \K2\ light curve (solid line) and $3\sigma$ contour level (dashed line). The x-axis of the Cepstrum is referred to as the ``quefrency'', and has unitis of time. Three significant peaks are detected at 1 hour, 4.4 hours and 12.6 hours. }
\label{fig:cep}
\end{figure*}

Finally we were also interested in determining whether the \K2\ optical lightcurve alone displays any echoes. Any detection of such features would be interesting for a number of reasons. For example, if the optical lag (response) detected in Fig. \ref{fig:DCF} is associated with the disk thermal timescale, we might expect an echo within the optical light curve alone around the same time. To determine whether any echoes are present in the \K2\ light curve we computed the so-called Cepstrum. The Cepstrum is computed as the inverse Fourier transform of the logarithm of the estimated spectrum of the light curve (see e.g. \citealt{bogert63,aframovic81,maccCep}), and is used to find repeating signals (echoes) within a single time-series. Confidence limits can also be computed with a similar method as that employed for the DCF confidence limits. We simulate $10^5$ light curves using the same input PSD as previously adopted, and determine for each timescale the $3\sigma$ level. The results of this is shown in Fig. \ref{fig:cep}. Three significant signals are detected at 1 hour, 4.4 hours and 12.6 hours. We will discuss possible mechanisms to generate these echoes in Section \ref{sec:discussion}, but already note here that the signal at 12.6 hours is consistent with the observed optical lag shown in Fig. \ref{fig:DCF}, and that the 4.4 hour signal is consistent with the FB DCF marginally detected secondary peak in Fig. \ref{fig:DCF2}. We also point out that the Cepstrum is solely based on the \K2\ light curve, thus any matching signals between the Cepstrum and the DCFs are unlikely to be coincidental.

\section{Discussion}\label{sec:discussion}

In this Section we discuss our findings in the context of the \cite{church12} model for Z-sources in general and Sco X-1 in particular. We also propose interpretations for several newly discovered phenomena.

\subsection{X-ray and optical state properties}
Phenomenologically, Sco X-1 transitions between NB and FB on timescales of $\approx1$ day. According to \cite{church12} the proposed mechanisms for generating these transitions is related to the mass-transfer rate varying, such that $\dot{M}$ increases along both the FB and NB tracks away from the vertex. The X-ray colour changes reflect the relative contribution to the X-ray flux of the black-body disc component (dominating in the soft X-ray band) and comptonised corona (dominating in the hard X-ray band). As the optical emission is thought to originate from a colder outer disk, the optical flux distribution better tracks $\dot{M}$ changes, being unaffected by X-ray absorption and/or the X-ray contribution by different X-ray emitting components. Thus, although both the X-ray and optical flux distributions appear bimodal, the bimodality found in X-rays may not correspond simply to states with different $\dot{M}$.

One important consequence of the \cite{church12} model (but also the older \citealt{psaltis95} model) is that as $\dot{M}$ increases, during the NB phase, the optical depth close to the NS also rises, due to higher radiation pressure. This might cause X-ray photons to be absorbed and re-emitted as optical photons, resulting in an anti-correlation between X-ray and optical fluxes. This is observed in Fig. \ref{fig:DCF2}. On the other hand, during the FB phase, the \cite{church12} model predicts an additional X-ray component caused by unstable nuclear burning on the NS surface. This additional X-ray component would shine on the optical emitting outer accretion disk, so we should observe a correlation between optical vs. X-ray fluxes. This is also observed as shown in Fig. \ref{fig:DCF2}. 

We have also found that the broad-band PSD shape of Sco X-1 differs substantially between the low and high optical luminosity states, with the latter displaying an additional broad noise component breaking at $\approx10^{-3}$ Hz superimposed $\approx-2$ power-law. \cite{bildsten93} has previously reported that a possible time-dependent accretion regime exists in accreting NS, where small patches of the NS burn intermittently (unstable and persistent, as opposed to burning steadily or rapidly on the entire NS surface). The fire will then propagate around the NS on the timescale it takes to accrete enough fuel for the next instability, which lies in the range between $10^{3}$ - $10^{4}$ seconds. Thus, low-level luminosity variations should be observed on a similar frequency range. Indeed \cite{hasinger89} have reported ``very low frequency'' noise observed on similar timescales. Given that the high optical luminosity state is somewhat associated to the FB for Sco X-1, and given that according to the \cite{church12} model this is where nuclear burning occurs on the NS surface, it is possible that the additional noise component in the high optical luminosity state observed here is in fact associated to the \cite{bildsten93} mechanism.       

It is also interesting to note how the rms-flux relations differ between the two low and high optical luminosity states. Linear rms-flux relations have been detected in all accreting compact objects (from Active Galactic Nuclei to accreting white dwarfs, see e.g \citealt{uttley05,heil12,scaringi12a,vdSande15}). The general model invoked to explain the linear rms-flux relations is that of the fluctuating accretion disk model (\citealt{lyub,kotov,AU06}). In this model, the observed variability is associated with modulations in the effective viscosity of the accretion flow at different radii. More specifically, the model assumes that, at each radius, the viscosity -- and hence accretion rate -- fluctuates on the local viscous timescale around the mean accretion rate, whose value is set by what is passed inwards (again on the viscous timescale) from larger radii. The overall variability of the accretion rate observed through light curve variations is therefore effectively the product of all the fluctuations produced at larger radii. It is important to realise, however, that the detection of the rms-flux relation only suggests that the observed variability is the result of multiplicative processes (rather than additive) and does not necessarily rule out more complex (or less complex) accretion disk models to generate the observed variability. 

In Sco X-1, the optical \K2\ light curve does always exhibit a positive correlation between mean flux and rms. However, this correlation is only linear in the low optical luminosity state. This suggests that during low optical luminosities the light curve is indeed the result of multiplicative processes, which are possibly the result of a fluctuating accretion disk. During times of high optical luminosity the rms-flux relation deviates from linearity (with an apparent change of slope to lower values) suggesting that more complex phenomena are at play during this phase. It is possible that the high optical luminosity state is composed of more than one accretion regime (for example the unstable nuclear burning mechanism of \citealt{bildsten93}), causing the higher fluxes of this phase to be non-stationary with respect to the lower fluxes. Whatever the reason, it is interesting to note that multiplicative processes alone cannot reproduce non-linear rms-flux relations. 

\subsection{Optical spectroscopy}
The \HERMES\ spectra reveal that the low ionisation lines of H$\alpha$, H$\beta$ and HeI appear stronger when Sco X-1 is in a low optical luminosity state (but not necessarily at low X-ray luminosities). \cite{fender09} have found a similar trend in a sample of XRBs, where an anti-correlation is found between the X-ray luminosity and the H$\alpha$ equivalent width. \cite{fender09} suggest that the observed anti-correlation is the result of optical-depth changes. 

In Sco X-1, the anti-correlation could be due to the outer disk becoming cold and optically thin in the continuum, but not the emission lines, during the transition between the two optical luminosity states (\citealt{williams80,fender09}). In this scenario the optical continuum would drop much faster than the ionizing continuum produced in the inner disc regions. The EW of the optical lines produced in the outer disc should thus increase. The same would also hold during the opposite transition from low to high states, as the continuum would rise much faster than the emission lines, thus swamping them out and decreasing their measured EWs. Another possibility is that during the high optical luminosities most of the accretion disk becomes too hot for the production of low ionisation emission lines. This would not necessarily cause higher ionisation lines such as HeII to decrease in EW. In this case optical fluxes alone (rather than X-ray colours) would provide a better tracer for $\dot{M}$ changes in the optical emitting region. Indeed 2 of our 6 spectra with the lowest optical fluxes display the highest EWs. Furthermore, we do not measure significant changes in the HeII EW, possibly making this scenario a more viable explanation. 

Although the HeII EW does not appear to significantly change between the low and high optical luminosity states, it does display a double-peaked profile only during low luminosity ones. We note that the HeII line profile changes do not appear to depend on the NB/FB dichotomy, but simply on optical luminosity (see Fig. \ref{fig:K2vsMAXI}). The change from double to single peaked profiles has been previously associated with the presence of outflows (see e.g. \citealt{murray96}). This deduction of a wind could potentially be consistent with the X-ray observations of the Iron $K_{\alpha}$ emission line modelling in Sco X-1 (\citealt{bradshaw07}), which associates a strong wind to an X-ray softening. 

\subsection{X-ray time-lags and optical echoes}
One other interesting feature is the broad $\approx12.5$ hour optical lag shown in Fig. \ref{fig:DCF}. Naively, from the fluctuating accretion disk model (\citealt{lyub,kotov,AU06}), we might have expected to observe an X-ray lag due to material propagating from an optical emitting region in the outer disk to an X-ray emitting region in the inner disk. Instead we observe the exact opposite for Sco X-1: first we observe X-rays and hours later the signal appears in the optical band. Given the hour-long measured timescales in the cross-correlation function of Fig. \ref{fig:DCF}, our result rules out light-travel time reverberation from the disk as a possible mechanism. We do not know at the time of writing what is the mechanism behind the observed optical lags but speculate that it could be the result of reprocessed X-ray photons to optical wavelengths on the disk thermal timescale. This idea is not new, and has been invoked to explain Fourier-dependent time-lags observed in accreting white dwarfs (\citealt{scaringi13}).

\begin{figure}
\includegraphics[width=0.5\textwidth]{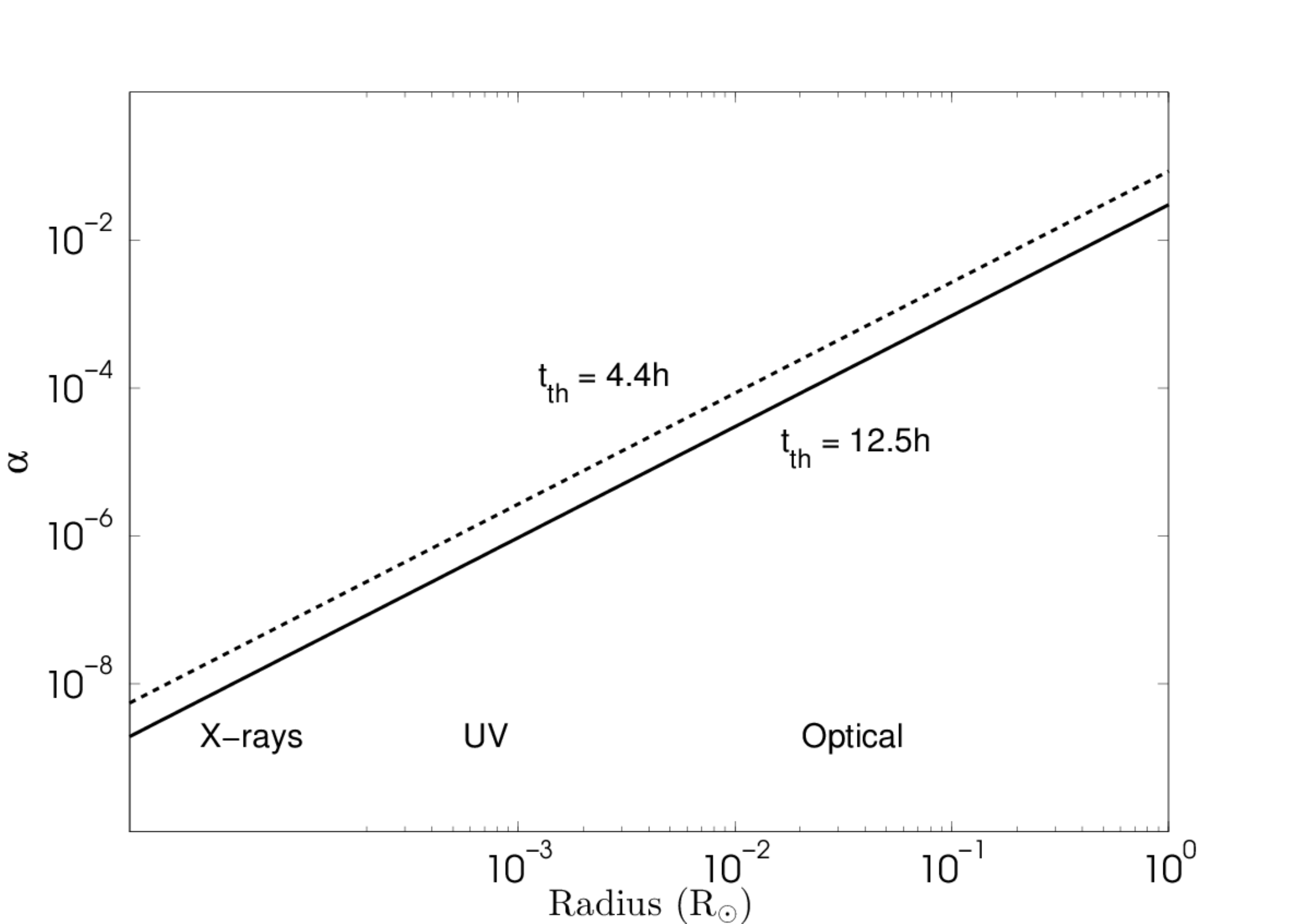}
\caption{$\alpha$ dependence with disk radius for $t_{th}\approx12.5$ hours (solid line) and $t_{th}\approx4.4$ hours (dashed line). We also mark approximate radii regions of different electromagnetic emission.}
\label{fig:alpha}
\end{figure}

Although the optical emission in the broad \Kepler\ pass-band will originate from a large radial disk component, we can attempt to place constraints on $\alpha$ since the thermal timescale $t_{th} = \frac{1}{\Omega\alpha}$ (where $\Omega$ is the dynamical frequency at some given radius in the disk and $\alpha$ is the viscosity parameter of \citealt {SS73}). Fig. \ref{fig:alpha} shows the obtained $\alpha$ values as a function of disk radius assuming a NS mass of $1.6M_{\odot}$ (\citealt{teo07}) by setting $t_{th}=12.5$ or $t_{th}=4.4$ hours (two of the peaks detected in the Cepstrum analysis). The constraints obtained from Fig. \ref{fig:alpha} are physically plausible, suggesting that the measured lag timescale could be associated to the disk thermal timescale. 

If the measured lags are really associated to the thermal timescale we would expect the X-ray and optical variability to have a very small, light-travel time, lag which we observe in Fig. \ref{fig:DCF} as the strong zero-lag peak. As some X-ray radiation will be absorbed by the disk and re-emitted on the thermal timescale at optical wavelengths we might then also expect a secondary, broad, optical lag peak. In this case we might expect the optical lightcurve to also display an echo, since the same signal should appear twice within the lightcurve (once reflected and once reprocessed), with a delay close to $t_{th}$. As one of the significant peaks found in the Cepstrum (Fig. \ref{fig:cep}) is similar to the measured broad X-ray lag, we tentatively associate this with the echo mechanism proposed above. A similar process might explain the other 2 detected echoes. In fact, the FB DCF in Fig. \ref{fig:DCF2} is qualitatively similar to the full DCF of Fig. \ref{fig:DCF}. Given our time reolution used to compute the DCFs, we would not be able to observe the 1 hour cepstral peak in the DCFs. We point out that our Cepstrum analysis is based on the full 78-day \K2\ light curve, and that we are not able at this time to localise the echoes within the light curve. Thus multiple features in the Cepstrum do not necessarily need to occur at the same time within the 78-day period.

We caution at this stage that our tentative association of the optical lags and Cepstrum signals to the disk thermal timescale are tentative, as they have never been observed on these timescales. However, given the long timescales involved, the non-detection of such lags in other similar systems is not surprising, since one would require continuous and long monitoring of XRBs at both X-ray and optical wavelengths. It is interesting to note the possibility that optical lags might only be detected in the FB state (since the NB shows X-ray/optical anti-correlation), and that the disk heating mechanism could be driven by the unstable nuclear burning on the NS surface. A detailed study into whether this scenario is feasible for Sco X-1 is however beyond the scope of this paper.

\section{Conclusion}\label{sec:conclusion}

A prime motivation for this study has been to revisit the well-known LMXB Sco X-1 with newly available data to test and refine the current physical model used to explain its phenomenological behaviour. Our dataset comprises a continuous 58.8 second cadence light curve over more than 78 days obtained with the \K2 mission, and represents the best ever optical light curve obtained for a LMXB. We additionally used X-ray data obtained with the \MAXI\ instrument on-board the ISS which provided us with X-ray light curves in the three energy bands 2-4 keV, 4-10 keV and 10-20 keV over the same observing period of \K2. Finally, we augmented our analysis with high resolution optical spectroscopy taken during the same period with the \HERMES\ instrument mounted on the 1.2 meter Mercator telescope. 

We can summarise the major findings of this paper as follows:

\begin{enumerate}
\item In the absence of X-ray colours, optical fluxes have a greater discriminatory power in isolating the NB and FB phases than X-ray fluxes alone. However, even optical fluxes alone cannot properly separate the two phases. 

\item The PSD of Sco X-1 appears to change in shape between low and high optical luminosity states. Qualitatively, the low-luminosity PSD can be expressed as a red noise PSD with a power-law $\approx-2$. During high optical luminosities an additional component is observed on top of the power-law with a possible break at $\approx10^{-3}$ Hz. This additional broad-band feature could possibly be associated to unstable nuclear burning on the NS surface (\citealt{bildsten93}).

\item Sco X-1 displays rms-flux relations in both the low and high optical luminosity states. However a linear relation is only observed during low luminosities. The deviation from linearity, during high luminosity states, might suggest that an additional variability component is present, thus removing the linearity and stationarity at the highest fluxes. This component could possibly be associated with the extra variability component observed in the FB PSD.  

\item We find that the equivalent widths of the low ionisation lines H$\alpha$ 6563 \AA, H$\beta$ 4861 \AA, HeI 5876 \AA, HeI 6678 \AA\ and HeI 7065 \AA\ are higher during times of lower optical luminosity, whilst HeII 4685 \AA\ does not appear to change in strength. This can be explained if during high optical luminosity periods (when $\dot{M}$ is highest) the accretion disc is too hot for the production of low ionisation lines, but not hot enough to remove higher ionisation lines such as HeII 4685 \AA. We also find that HeII 4685 \AA\ appears to be double-peaked only during high optical luminosity states. It is possible that the single-peaked profile of HeII 4685 \AA\ is associated to an accretion disk outflow during the low optical luminosity phase. This interpretation is in line with previous X-ray measurements of the Iron $K_{\alpha}$ emission line model (\citealt{bradshaw07}), where fast outflows are supposed to be generated in Sco X-1 during periods of high mass transfer.

\item The DCF between the available optical \K2\ and X-ray \MAXI\ lightcurves displays clear asymmetry. The DCF profile displays a clear, broad, 12.5 hour secondary optical lag peak. This can potentially be attributed to the disk thermal reprocessing timescale.

\item The DCF obtained during the FB also displays clear asymmetry, and is qualitatively similar to the full DCF, with a secondary marginal detection of a 4.5 hour optical lag peak. This too could potentially be the signature of the disk thermal timescale. It is thus possible that the broad optical lags are mostly generated during the FB state.

\item The DCF obtained during the NB clearly show that the optical and X-ray variability are anti-correlated during this phase. This phenomenon has been previously reported by other authors. 

\item We detect clear echoes within the optical \K2\ light curve alone using the Cepstrum, at 1 hour, 4.4 hours and 12.6 hours. The 12.6 hour echo timescale is consistent with the full optical/X-ray DCF, whilst the 4.4 hour echo is consistent with the FB DCF. Even if it were present in any of the DCFs studied here, we would not be able to recover the 1 hour echo due to our DCF resolution. We interpret these echoes as the disk thermal timescale, connected to the DCF optical lags. Thus, as some of the X-ray radiation will be reflected quasi-instantaneously by the outer optical emitting disk, some radiation will be absorbed and re-emitted on the thermal timescale. This would give rise to the broad optical lags observed. The reprocessing mechanism would also generate echoes within the optical lightcurve alone, with a delay close to the thermal timescale.

\end{enumerate}

\section*{Acknowledgements}
S.S. acknowledges funding from the Alexander von Humboldt Foundation. G.P. acknowledges support by the Bundesministerium f{\"u}r Wirtschaft und Technologie/Deutsches Zentrum f{\"u}r Luft- und Raumfahrt (BMWI/DLR, FKZ 50 OR 1408) and the Max Planck Society. The authors also acknowledge Saul Rappaport and Daryll LaCourse for useful discussions relationg to the \K2\ lightcurve, as well as the anonymous referee for useful comments which have imporved the manuscript. This research has made use of NASA's Astrophysics Data System Bibliographic Services. Additionally this work acknowledges the use of the astronomy \& astrophysics package for Matlab (\citealt{matlab}). This paper includes data collected by the Kepler mission. Funding for the Kepler mission is provided by the NASA Science Mission directorate. Some of the data presented in this paper were obtained from the Mikulski Archive for Space Telescopes (MAST). STScI is operated by the Association of Universities for Research in Astronomy, Inc., under NASA contract NAS5-26555. Support for MAST for non-HST data is provided by the NASA Office of Space Science via grant NNX13AC07G and by other grants and contracts. This research has made use of the MAXI data provided by RIKEN, JAXA and the MAXI team. This work is also based on observations obtained with the HERMES spectrograph, which is supported by the Fund for Scientific Research of Flanders (FWO), Belgium, the Research Council of K.U.Leuven, Belgium, the Fonds National de la Recherche Scientifique (F.R.S.-FNRS), Belgium, the Royal Observatory of Belgium, the Observatoire de Genève, Switzerland and the Thüringer Landessternwarte Tautenburg, Germany.

\bibliographystyle{mn2e}
\bibliography{scox1_paper}

\end{document}